\documentclass[sigconf]{acmart}
\usepackage{xcolor}
\usepackage{hyperref}
\usepackage{subcaption}
\usepackage{graphicx}
\usepackage{xspace}
\usepackage{listings}

\usepackage{tabularx}
\usepackage{multirow}

\definecolor{light-gray}{gray}{0.9}
\definecolor{dark-gray}{gray}{0.6}

\AtBeginDocument{%
  }

\copyrightyear{2024}
\acmYear{2024}
\setcopyright{rightsretained}
\acmConference[ICSE-Companion '24]{2024 IEEE/ACM 46th International Conference on Software Engineering: Companion Proceedings}{April 14--20, 2024}{Lisbon, Portugal}
\acmBooktitle{2024 IEEE/ACM 46th International Conference on Software Engineering: Companion Proceedings (ICSE-Companion '24), April 14--20, 2024, Lisbon, Portugal}\acmDOI{10.1145/3639478.3643122}
\acmISBN{979-8-4007-0502-1/24/04}

\begin{document}

\makeatletter
\def\@ACM@checkaffil{%
    \if@ACM@instpresent\else
    \ClassWarningNoLine{\@classname}{No institution present for an affiliation}%
    \fi
    \if@ACM@citypresent\else
    \ClassWarningNoLine{\@classname}{No city present for an affiliation}%
    \fi
    \if@ACM@countrypresent\else
        \ClassWarningNoLine{\@classname}{No country present for an affiliation}%
    \fi
}
\makeatother

\title{Assessing AI-Based Code Assistants in Method Generation Tasks}

 \author{Vincenzo Corso, Leonardo Mariani, Daniela Micucci and Oliviero Riganelli}

 \affiliation{ \institution{University of Milano-Bicocca}
 \city{Milan}
  \country{Italy}}

\newcommand{\LEO}[1]{\textcolor{blue}{{\it [Leonardo says: #1]}}}
\newcommand{\DAN}[1]{\textcolor{red}{{\it [Daniela says: #1]}}}

\newcommand{\oli}[1]{\textcolor{red}{#1}} 

\newcommand{\urlRepo}[0]{\url{https://shorturl.at/krBK7}\xspace}

\begin{abstract}
AI-based code assistants are increasingly popular as a means to enhance productivity and improve code quality. This study compares four AI-based code assistants, GitHub Copilot, Tabnine, ChatGPT, and Google Bard, in method generation tasks, assessing their ability to produce accurate, correct, and efficient code. Results show that code assistants are useful, with complementary capabilities, although they rarely generate ready-to-use correct code.%
\end{abstract}

\maketitle

\section{Introduction}

\noindent \textbf{Context:} AI-based code assistants are becoming increasingly popular. For instance, recent studies demonstrated that they can provide useful code snippets~\cite{Vaithilingam22,Yetistiren:PROMISE:22,Dakhel:JSS:2023}. However, they are still limited in the generation of code ready-to-be integrated into real-world programs.

\noindent \textbf{Problem:} To address these limitations, this study compares four AI-based code assistants - Copilot~\cite{Copilot2023}, Tabnine~\cite{Tabnine2023}, ChatGPT~\cite{ChatGPT2023}, and Bard~\cite{GoogleBard2023} - on their ability to generate code for 100 Java methods extracted from real-world open source projects.

\noindent \textbf{Methodology:} The study executes the four assistants using the comment and the signature associated with the selected methods as prompts. The quality of the generated code is evaluated according to five criteria: functional correctness, complexity, efficiency, size, and similarity to the original code produced by developers.

\noindent \textbf{Main results:} Copilot emerged as the most effective assistant in this task, although all assistants demonstrated their strengths. The study revealed the need for improvement in handling inter-class dependencies. Surprisingly, the generated code sometimes outperforms developer-written code.

\noindent \textbf{Extended Abstract:} This extended abstract summarizes results extensively presented at the International Conference on Program Comprehension~\cite{Corso:ICPC:2024}. The full paper describes the methodology, the results, and the findings more in detail. %
\vspace{-1mm}
\section{Methodology}
In this study, we assess AI-base code assistants by investigating
five research questions:

\noindent \textbf{RQ1 - Is the code generated by AI-based code assistants \emph{correct}?} This RQ investigates the syntactic and semantic correctness of the generated code.

\noindent \textbf{RQ2 - What is the \emph{McCabe complexity} of the generated code?} This RQ investigates if AI-based code assistants cannot only generate correct code but also produce code with a level of complexity similar to the code implemented by developers.

\noindent \textbf{RQ3 - How  \emph{efficient} is the generated code?} This RQ investigates if the generated correct code is as efficient as the one implemented by developers.

\noindent \textbf{RQ4 - What is the \emph{size} of the generated code?} This RQ investigates if the size of the generated code is similar to the code implemented by the developers.

\noindent \textbf{RQ5 - How \emph{far} is the generated code from the one implemented by developers?} This RQ studies the similarity of the code implemented by the developers to the code generated by the experimented tools, according to change-oriented static metrics.

The methodology for this study involves a multi-step process designed to address each research question systematically:

\begin{enumerate}
\item \textbf{Dataset Construction:} Selection of 100 Java methods from well-ranked GitHub projects, ensuring diversity in complexity and relevance. %
    
\item \textbf{Code Generation:} Employing four prominent AI-based code assistants - Copilot, Tabnine, ChatGPT, and Bard - to generate code for the selected methods, using both the method-level comment and the signature as prompt.
    
\item \textbf{Code Evaluation:} Assessing the generated code against the developer-implemented code in terms of correctness, complexity, efficiency, size, and similarity.
    
\item \textbf{Statistical Analysis:} Using statistical methods to compare the performance of AI-based code assistants and identify significant differences.
\end{enumerate}

To ensure a realistic and comprehensive evaluation of AI-based code assistants, the study constructs a dataset of real-world Java programming tasks. The dataset is designed to include methods of different levels of complexity and with different types of dependencies, including stand-alone methods, methods with intra-class dependencies, and methods with inter-class dependencies. %
By restricting the selection to methods that appeared in recent GitHub commits, we mitigated the risk of using code that had been considered during the training of the assistants that we assessed. 

\section{Results}

\noindent \textbf{RQ1 - Code Correctness.} Copilot generated the highest percentage of correct methods, achieving a 32\% success rate. ChatGPT followed with 23\% correct methods generated. Finally,  Bard and Tabnine achieved 15\% and 13\% success rates, respectively. %

Even if each assistant demonstrated some unique capabilities, all assistants can still be largely improved. In fact, a non-negligible portion of the generated code was incorrect, especially when the method requires dealing with inter-class dependencies, where the best-performing assistant, Copilot, achieved only 15\% correctness. %
 
 We also noticed a remarkable difference between correct code (i.e., code equivalent to developers' code based on human inspection) and plausible code (i.e., code that passes the available test cases). In our experiment, assistants generated 31\% plausible methods, but only 21\% correct methods on average, confirming that tests cannot accurately establish the correctness of the generated code.

\noindent \textbf{RQ2 - Code Complexity.}
The four assistants generally generated code with similar McCabe complexity to the code written by developers. In some cases, the generated code had slightly higher complexity due to the use of explicit if conditions, or lower complexity due to the use of lambda expressions and methods that encapsulate checks. Overall, the generated code was of similar complexity to the original. The four tools did not differ significantly in terms of the complexity of the generated code.

\noindent \textbf{RQ3 - Code Efficiency.} 
The four assistants generated code that is as efficient as, or more efficient than, the code written by developers. A significant portion of the generated methods, 87\% for ChatGTP and 100\% for Tabnine, exhibited no significant difference in execution time compared to the original methods. In some cases, the generated code even outperformed the original code. The exceptions were Copilot and ChatGPT, which generated a small number of methods that were slower than the original ones. These inefficiencies were attributed to suboptimal data type choices, unnecessary operations, inefficient control flow, and redundant method calls.

\noindent \textbf{RQ4 - Code Size.} 
We compared the number of lines of code (LOCs) in the generated code and the code written by the developers for all the methods in our dataset. The results show that the size of the generated and original code is similar and that the four code generation tools tend to generate code of similar length.  ChatGPT and Bard produced code with the highest difference and variance in length compared to the length of the code written by developers. %

\noindent \textbf{RQ5 - Code Similarity.} 
The four assistants produced code that is significantly different from the code written by the developers, with similarity measured according to the normalized Levenshtein similarity~\cite{Levenshteinsw} and CodeBLEU scores~\cite{CodeXGLUE}. For the incorrectly generated code, the distances from the developers' code are larger than the correct code. Tabnine generated the correct code that is most similar to the developers' code (median CodeBLEU of 0.528). 

This result suggests that although the generated code could be close to the intended code in complexity and size, it still has to be significantly adjusted to fully match the expected implementation.

\medskip

The results indicate that AI-based code assistants can be a valuable tool for developers, but they also need to be improved. While the assistants can generate code that is generally correct and efficient, they also produce a significant amount of invalid or incorrect code, particularly for methods with inter-class dependencies. Furthermore, the generated code often differs significantly from developer-written code, requiring substantial revisions. These findings highlight the need for further development and refinement of AI-based code assistants to enhance their accuracy, efficiency, maintainability, and resemblance to developer-written code.

\section{Implications and Conclusions}

The study's findings have several implications for both research and practice. First, the collaboration among multiple AI-based code assistants is a promising research direction. Developers should consider leveraging the strengths of different assistants to enhance code generation and address the limitations of individual assistants.

 Second, the study revealed that AI-based code assistants can sometimes generate better code than the code implemented by developers. This finding has implications for improving code quality and efficiency in software development.

 Third, the challenges with inter-class dependencies highlight the need for further improvement in AI-based code assistants to effectively handle dependencies that extend beyond the boundaries of single classes. Future research and development efforts should focus on enhancing the capabilities of AI-based tools to address complex inter-class dependencies in code generation.

In conclusion, AI-based code assistants have the potential to significantly improve code generation and quality, but further research is needed to address existing challenges and limitations. This study investigates the capabilities of four AI-based code assistants: GitHub Copilot, Tabnine, ChatGPT, and Google Bard. Their capabilities are compared according to the functional correctness, complexity, efficiency, size, and similarity to the original code. %

Assistants demonstrated to have complemental capabilities, with Copilot generating the highest rate of correct methods. Results also reveal that the generated code could be a good starting point to derive the actual implementation, but it seldom consists of ready-to-use code. The capability to deal with inter-class dependencies is recognized as one of the main limitations.

\begin{acks}
This work has been partially supported by the Engineered MachinE Learning-intensive IoT systems (EMELIOT) national research project (PRIN 2020 program Contract 2020W3A5FY).
\end{acks}

\vfill

\balance
\bibliographystyle{ACM-Reference-Format}

%
%
\end{document}